\documentclass[preprintnumbers,amsmath,amssymb,twocolumn,showpacs]{revtex4}
\usepackage{graphicx}
\usepackage{dcolumn}
\usepackage{bm}

\begin{document}

\title{Spin Wave Diffraction Control and Read-out with a Quantum Memory for Light}

\author{Gabriel H\'etet$^{1}$} 
\author{David  Gu\'ery-Odelin$^{2,3}$}
\affiliation{$^1$ Laboratoire Pierre Aigrain, Ecole Normale Sup\'erieure-PSL Research University, CNRS, Universit\'e Pierre et Marie Curie-Sorbonne Universit\'es, Universit\'e Paris Diderot-Sorbonne Paris Cit\'e, 24 rue Lhomond, 75231 Paris Cedex 05, France}
\affiliation{$^2$ Universit\'e de Toulouse ; UPS ; Laboratoire Collisions Agr\'egats R\'eactivit\'e, IRSAMC ; F-31062 Toulouse, France} 
\affiliation{$^3$ CNRS ; UMR 5589 ; F-31062 Toulouse, France}

\begin{abstract}
A scheme for control and read-out of diffracted spins waves to propagating light fields is presented.
Diffraction is obtained via sinusoidally varying lights shifts and ideal one-to-one mapping to light is realized using a gradient echo quantum memory. 
We also show that dynamical control of the diffracted spin waves spatial orders can be implemented to realize a quantum pulse sequencer for temporal modes that have high time-bandwidth products. 
Full numerical solutions suggest that both co-propagating and couterpropagating light shift geometries can be used, making the proposal applicable to hot and cold atomic vapours as well as solid state systems with two-level atoms.
\end{abstract}
%
\maketitle

Macroscopic spin superpositions in atomic ensembles have been studied for decades and have found use in many areas of physics. 
They are essential for quantum optical information storage and retrieval \cite{Lvovsky} and for sensitive magnetometry \cite{Budker}. 
Spin superpositions containing quantized excitations are prepared either by performing quantum non-demolition measurements on the scattered light \cite{DLCZ, Kuzmich, SewellR, Haas27032014}, or by directly mapping quantum states of light \cite{Eisaman, Hald, Appel, Honda} onto the atoms. Using appropriate schemes furthermore enables the actual shape of a spin wave to be controlled to a high degree.
Andr\'e {\it et al.}  \cite{andr} proposed a scheme to perform Bragg scattering of admixtures of spin and light components with counterpropagating lasers and electromagnetically induced transparency (EIT). Two counterpropagating polariton modes are coupled, enabling light fields to be localised within the medium with far reaching prospects in quantum non-linear optics \cite{AndrNL} and quantum simulations \cite{Otterbach, Otterbach2}. A signature of polaritonic Bragg diffraction was observed in the fluorescence induced by the stationary light \cite{Baj03} but observing the diffracted modes is currently challenging due to the drastic requirements on the atomic ensemble temperature and the lack of direct read-out procedure. 
These methods are also not applicable for two-level atom based quantum memories \cite{Hed10, Bonarota}. 

\begin{figure}[ht!]
\centerline{\scalebox{0.2}{\includegraphics{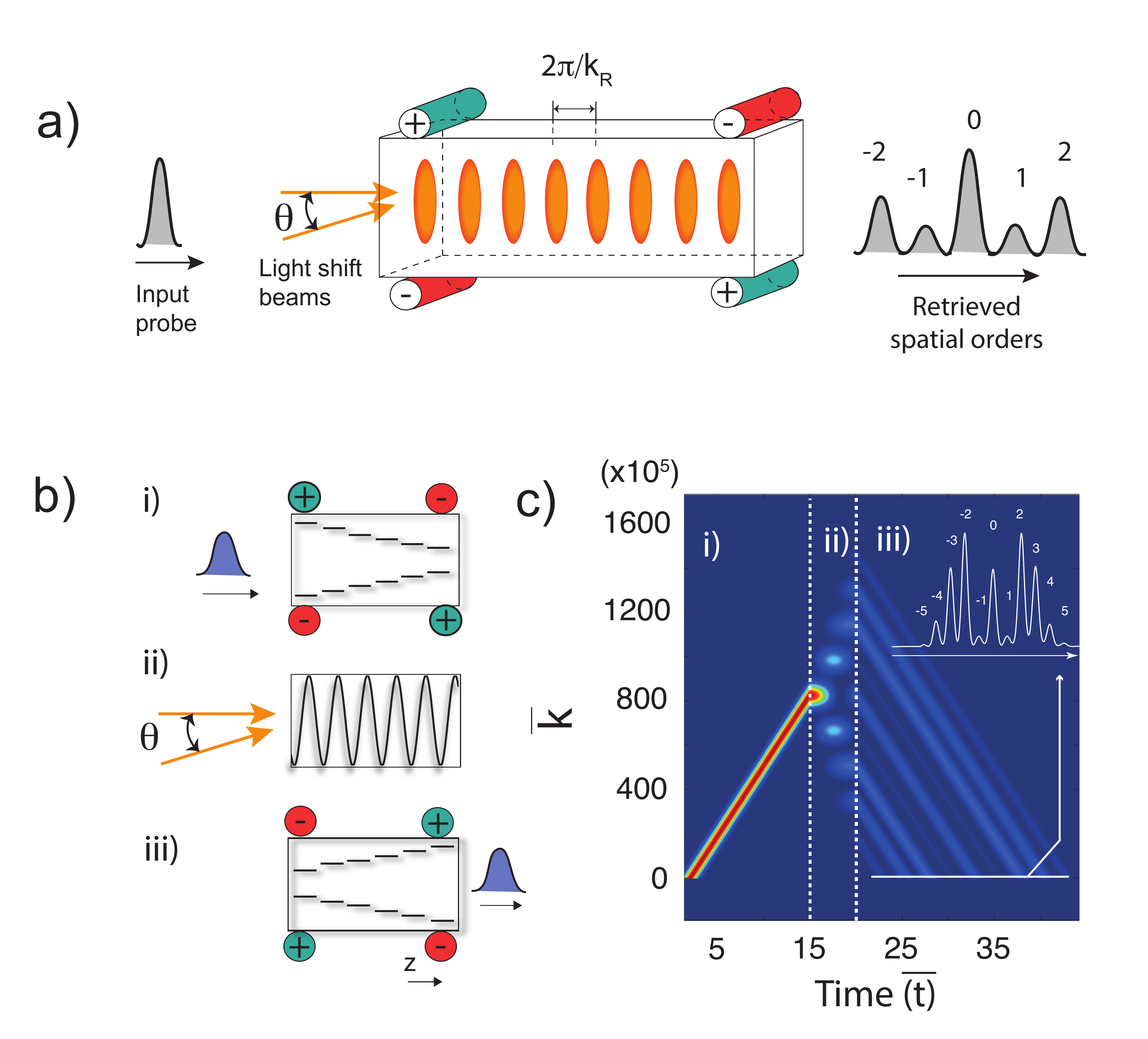}}}
\caption{a) Gradient echo memory (GEM) scheme with a longitudinal diffraction grating applied during storage. b) i) Creation of a narrow $k$-space distribution of spin waves by mapping a pulse of light in the memory. ii) Diffraction of the polariton by applying a pair of intense detuned lasers with an angle $\theta$ for a time $\tau$. iii) The detuned lasers are turned off and the read-out of the diffraction pattern is done by flipping the gradient sign. 
c) 2D-map of the atomic coherence amplitude in $k$-space as a function of time, demonstrating Kapitza-Dirac diffraction of the spin coherence wave.}\end{figure}

An alternative for realising Bragg scattering of spin waves is through the use of a controlled inhomogeneous broadening, and a prominant technique is the gradient echo memory (GEM) \cite{alexander, Het208}.
GEM was already shown to allow light storage at the quantum level with two \cite{Hed10} and three atomic levels \cite{Hos11}, to demonstrate efficient pulse re-ordering \cite{Hos2009} and spectral manipulation of light pulses that have large time-bandwidth products \cite{Sparkes}.
Here, we demonstrate that GEM can be used in conjunction with a grating to enable efficient coherent mapping between diffracted modes and propagating light fields.
For this, we use a theory that includes all spatial orders in scattering theory and show that one-to-one mapping is possible even in a room temperature vapour cell, beyond the secular approximation \cite{Nikoghosyan}. Last, we show that for specific durations of the grating imprint, the coupling of the spin wave to the output field is efficiently canceled and that diffraction can actually be temporally reversed by displacing the grating nodes. These two effects together realize a two-level atom coherent pulse sequencer for light pulses that have large time-bandwidth products. 

To perform longitudinal diffraction of a spin wave one
 creates a spin wave with longitudinal momentum spread $\Delta k$ by writing a weak laser field excitation in an optically dense atomic ensemble.
Using GEM, an effective two-level atomic ensemble is first subjected to a controlled linearly varying gradient $\eta(t)z$ as depicted in Figure 1.
After an input pulse with temporal width $\Delta t$ enters the medium, polaritonic modes with increasing momentum $k$ over time are created. When the gradient is turned off, polaritons remain trapped in the ensemble with a longitudinal momentum extension $\Delta k=\eta \Delta t $.
Here, we suppose that the effective two-level atoms form a quasi one-dimensional ensemble along a direction $z$. The corresponding operators are $\sigma_{12}^j(z,t_0)$, where $|1\rangle$ and $|2\rangle$ are the two atomic states and $j$ labels atoms within a slice with length $\delta z$.
The locally averaged coherence operators read
$\sigma_{12}(z,t_0)=\frac{1}{N_z}\sum_{j=1}^{N_z} \sigma_{12}^j (z,t_0)$,
where $N_z$ is the number of atoms in a volume given by the cross section area of the interacting light field over a slice $\delta z$.
We now suppose that the atomic superpositions do not dephase over the course of the experiment. In particular, this means that we can ignore the Langevin operators and treat the operators $\sigma_{12}$ as c-numbers.

To realize diffraction of the spin wave, a pair of intense pulses with an angle difference $\theta$, with a Rabi frequency $\Omega$ and detuning $\Delta$ from the two-level atomic transition are then applied to the atomic sample prepared in a spin superposition. 
If the detuning is larger than the decay rate from the excited state, the atoms will thus interact with a standing wave that produces a sinusoidally varying light shift $\delta(z,t)=\nu(t)\cos(k_R z)$ on the atomic coherences, where $\nu(t)=\Omega(t)^2/\Delta$. 
$k_R$ is related to the angle between the two lasers that generate the optical grating and must be greater than $\Delta k$
\footnote{For instance, if the laser fields are quasi co-propagating, a slight angle $\theta\approx 1$ degrees between them yields a 
fringe spacing $d=2\pi/k_R=2\lambda/\theta^2$, which, with laser fields in the visible range, can yield grating periods ranging from hundreds of micrometers to centimeters.}.
In rescaled time and space units (normalized by $l=c/g_0$ and $1/g_0$ respectively where $g_0$ is the vacuum Rabi frequency in the light mode and $c$ the vacuum speed of light), the Maxwell-Bloch equations read:
\begin{eqnarray}\nonumber
\frac{\partial}{\partial \overline{z}} \mathcal{E} &=& i \sqrt{N} \sigma_{12}\\ \label{MB}
\frac{\partial}{\partial \overline{t}} \sigma_{12} &=& i \Big[ \overline{\eta}(\overline{t})\overline{z}+\overline{\nu}(\overline{t}) \cos(\overline{k}_R\overline{z})\Big] \sigma_{12} +i \sqrt{N}\mathcal{E}, 
\end{eqnarray}
with $\overline{\eta}=\eta l/g_0$, $\overline{\nu}=\nu/g_0$ and $N$ is the atomic density.
Transforming these equations into the spatial Fourier domain, one gets an evolution equation for the spin wave inside the sample \cite{Het08}
\begin{eqnarray}\label{EQpolr}\nonumber
\Big(\frac{\partial}{\partial \overline{t}} -\overline{\eta}(\overline{t}) \frac{\partial}{\partial \overline{k}} +i\frac{N}{\overline{k}} \Big  ) \Psi(\overline{k},\overline{t})\\
= \overline{\nu}(\overline{t})\big[\Psi(\overline{k}+\overline{k}_R,\overline{t})&+&\Psi(\overline{k}-\overline{k}_R,\overline{t})\big],
\end{eqnarray}
where $\Psi=\overline{k} \mathcal{E}+ N \sigma_{12}$.
In this picture, absorption or emission of light occurs when the polariton reaches $\overline{k}=0$, i.e. when the phase matching condition is met.

Let us now consider the time-of-flight-like sequence depicted in Fig.~1-b). i) A pulse of light is sent to the sample 
at $\overline{t}_{\rm in}=1.5$ with a positive gradient slope and with $\overline{\nu}=0$. 
ii) The gradient is turned off and the grating turned on at a time $\overline{t}_0=15$. Assuming that the optical field is fully transfered to the atoms, applying the grating to the atoms for a time $\tau$ yields the following polariton evolution
\begin{eqnarray}
\Psi(k,t_0+\tau)
&=& \sum_{n=-\infty}^{+\infty} i^n J_n(\nu \tau) \Psi(k+n k_R,t_0),\label{KD}
\end{eqnarray}
where $L$ is the sample length and $J_n$ is the $n$-th order Bessel function of the first kind.
We thus expect the initial wave to be splitted into multiple spatial orders with a weight given by the Bessel functions. 
iii) The grating is finally turned off at a time $\overline{t}=20$ and the gradient is turned on with a negative slope to complete the gradient echo protocol.
The result of numerical simulations is displayed on the right panel of Figure 1-c), showing the evolution of $ |\Psi(k,t)|^2$ as a function of rescaled space and time. For the grating parameters, we choose $\overline{\nu}=\nu/g_0=2$, $\overline{k}_R=k_R l=150\times10^{5}$ which, with $g_0=3$ kHz, corresponds to a grating step of 16~mm. We also use $\overline{\eta}=7.10^6$ and an optical depth $2\pi N^2/\overline{\eta}=5.4$. 
Numerical simulations show that in the mapping phase the polariton evolves to higher $k$ until the gradient is turned off. When the grating is turned on Bragg diffraction is clearly observed.
The zeroth order mode is depleted and higher order modes are populated as a function of time, as expected from Eq.~(\ref{KD}).
The light-shift $\delta(z,t)$ thus leads to an evolution of the collective spin wave that is analogous to the Kapitza-Dirac diffraction \cite{Freimund}. 

At $\overline{t}=20$ the grating is turned off and the gradient turned on. As expected, the light field is reemitted at $k=0$.
One specific feature of GEM here is that, as is manifest in Fig.~1-c), the energy of the spin wave modes is released from the memory at times $\overline{T}=\overline{T_0}+n \overline{k_R}/\overline{\eta}$, where $n$ denotes the mode order and $T_0$ the time at which the zeroth order mode comes out.
The polariton spatial modes are here in the positive range before switching so they are all released from the memory with a period $\overline{T}=\overline{k_R}/\overline{\eta}=2.2$. In typical diffraction grating experiments with light, electrons \cite{Freimund}, degenerate atomic gases \cite{Fabre, Martin} or molecules \cite{Gerlich},  
the field onto which a periodic phase profile is imprinted is let free to expand in two-dimensions to realize the equivalent of a Fourier transformation. This operation maps the local phase onto a transverse profile that displays the discrete set of spatial modes. We demonstrated here that GEM exactly realizes such a Fourier transform but in the time domain, allowing efficient read-out of the diffracted longitudinal modes in this time-of-flight like sequence.

\begin{figure}[ht!]
\centerline{\scalebox{0.20}{\includegraphics{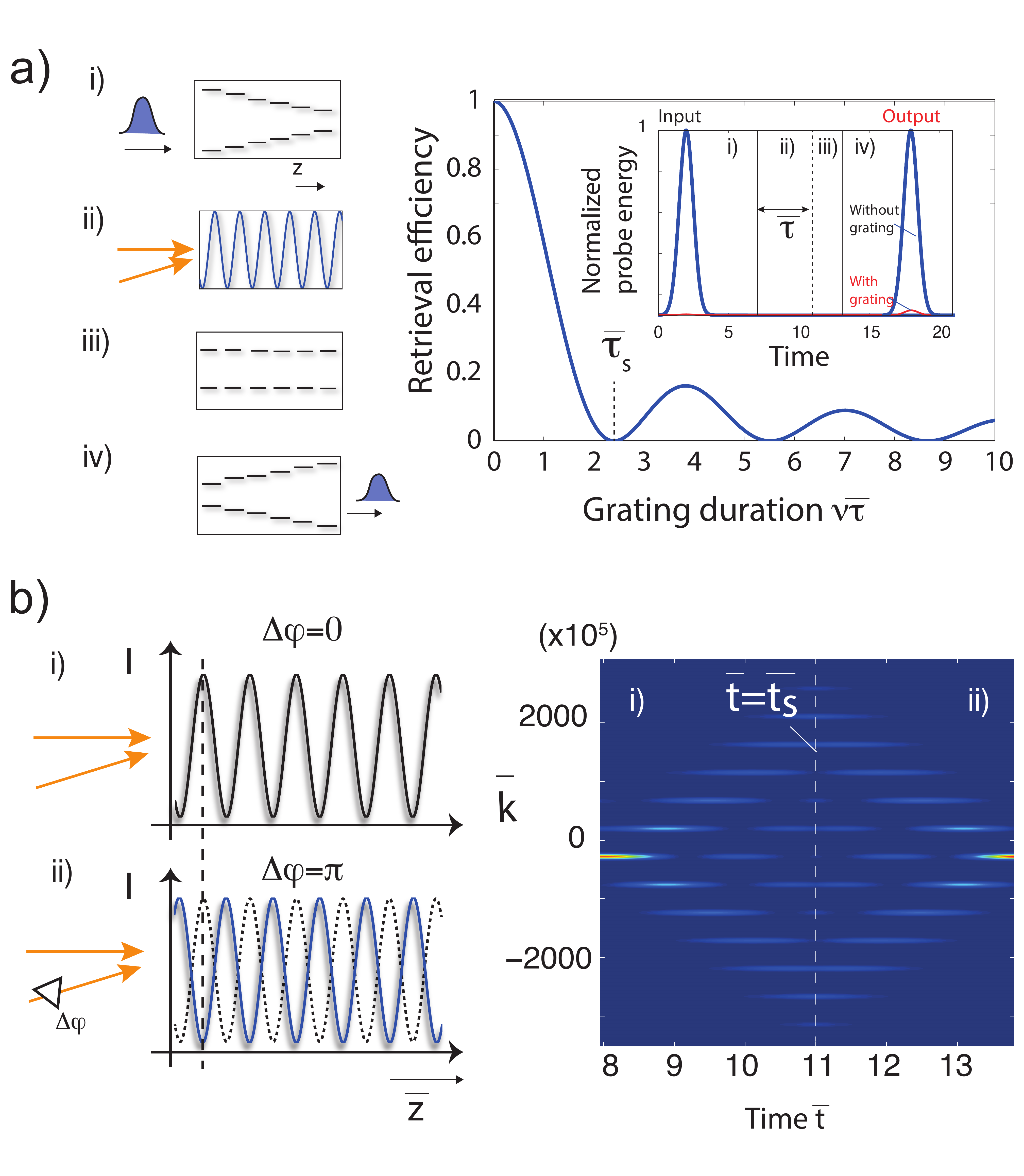}}}
\caption{a) Simulations of the gradient echo memory (GEM) scheme with a diffraction grating applied during storage. Light read-out efficiency as a function of grating duration $\tau$. Inset : Input and output probes with and without the grating.
b) Refocussing of the diffraction pattern. At a time $\overline{t}=\overline{t_s}$, the grating position is displaced by $\lambda/2$ by shifting one of the arms phase by $\pi$ (left panel), producing rephasing of the diffraction pattern to the zeroth order spatial mode (right panel).
}\end{figure}

We now turn to the demonstration of a quantum pulse sequencer.
At this point, we shall distinguish two operating regimes for diffraction read-out: the regime I where $n' k_R > k_0=\eta (t_0-t_{\rm in})$ and regime II where $n' k_R < \eta (t_0-t_{\rm in})$, where $n'=\lceil \nu \tau \rceil$ is the number of populated modes for a grating duration $\tau$.
In regime I, $n'$ spatial modes exactly occupy both the positive and negative $k$-values.
In regime II, $n'$ spatial modes only occupy the positive $k$-space, as per Figure 1-b).

Simulations are now realised with the sequence shown in the inset Figure~2-a) and in the regime I, reached by simply increasing the grating momentum by a factor of 3.3 and reducing the gradient by the same factor.
Figure~2-a) shows the retrieval efficiency of the zeroth order mode as a function of grating duration in a time window $\overline{t}=18\pm 3$. As the duration of the interaction with the grating is increased, the zeroth order mode gets depleted leading to a drop in retrieval efficiency. This is because the polariton keeps the energy in higher order spatial modes that remain trapped in the atomic ensemble. The first positive spatial mode will only be phase matched to the optical field a time $k_R/\eta$ after the zeroth order mode is emitted. The energy of the zeroth order mode released from the memory follows the zeroth order Bessel function, meaning that the energy is coherently sloshing back and forth between stationary spatial modes inside the ensemble before being output, a signature of the wave nature of the spin coherences. Remarkably, all the energy remains trapped inside the sample at a time $\tau_s$=2.4 corresponding to the first node of the zeroth order Bessel function. As we will show, this offers a new useful degree of freedom for sequencing light pulses. 

Another important ingredient for pulse reordering is refocussing of the spatial modes. The presented diffraction originates from a superposition of the electron dipole amplitudes of stationary atoms. It is therefore in principle reversible provided they stay in a pure state throughout the coupling to the grating.
Spatial mode refocussing can be achieved quite simply using a spin-echo like protocol, as shown figure 2-b) : following diffraction, the phase of one of the arms of the grating is shifted by $\pi$ at a time $\overline{t}_s=11$ to exactly cancel the phase acquired by each coherence. This can readily be done experimentally using acousto-optic modulators. 
As shown Fig. 2-b) (right panel) the atomic coherence evolution indeed refocusses so that the zeroth order spatial mode is recovered with unit efficiency, demonstrating the reversibility of the diffraction process. 

The effects presented Fig 2 can also be observed using other memory protocols. For instance with homogeneous three-level $\Lambda$ atomic memory schemes such as EIT \cite{Gor07} or Raman coupling \cite{Reim}, the grating can be turned on when the control fields are off, that is when the optical field is transferred to the atomic coherences. The GEM conveyor belt in $k$-space is however essential for diffraction imaging.
Furthermore, the last two properties of the presented spin wave diffraction have direct implications for precision manipulation of optical pulses and single photons using GEM.
It was already shown that GEM enables efficient storage of pulses that have large time-bandwidth products without the requirement for high optical depths \cite{Het08}. 
Using a Raman transition and a time dependent coupling to the ground state, pulse re-ordering was also demonstrated \cite{Hos2009}. For instance, input pulses were shown to come out first-in-first-out (FIFO) by turning off the coupling to the ground state at the first $k=0$ crossing and by switching the gradient twice using a $\Lambda$-GEM. 
Extending these effects to two-level atomic memories is important as it would enlarge the range of material systems where random access memories can be implemented. A combination of diffraction intensity and refocussing control in the regime I actually provides the ingredients to do so very efficiently as we now show.

\begin{figure}[t!]
\centerline{\scalebox{0.22}{\includegraphics{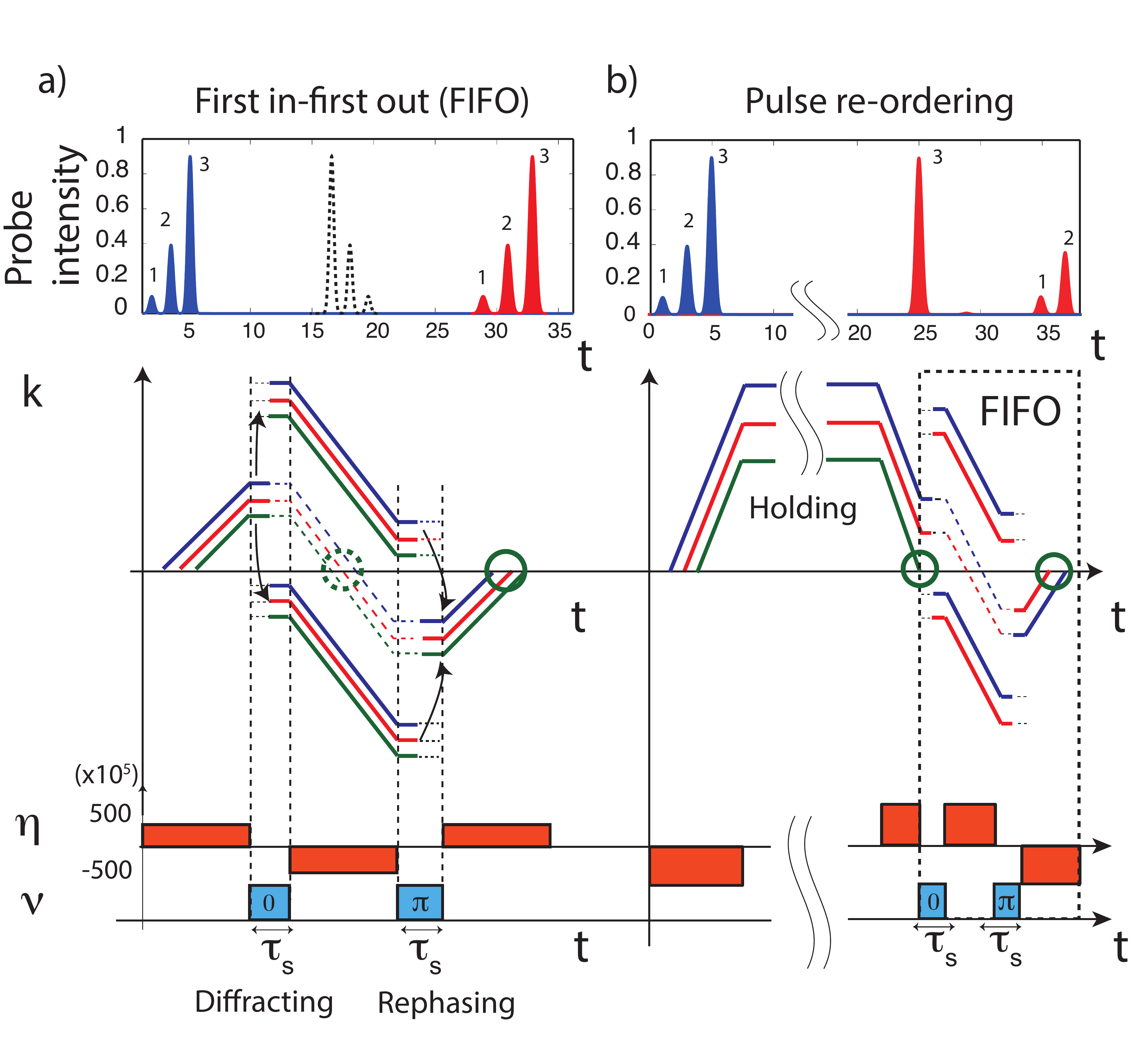}}}
\caption{Re-ordering of pulses using two level atoms. From top to bottom : probe intensity as a function of time, evolution of the energy of the polariton in k-space and gradient slope and grating amplitudes used in the numerical simulations. a) First in first out scheme. b) Pulse reordering.}\end{figure}

The idea is shown Fig. 3-a). Three pulses are input to the GEM memory. At $t=10$, one turns on a grating on for a time $\tau_s$ such that the zeroth order mode is completely depleted, that is at the first node of the zeroth order Bessel function. The gradient is then reversed. Since the zeroth order mode is depleted, when the $k=0$ condition is reached, there is no emission of light and all the pulse energy is mapped onto higher order spatial modes. One can however retrieve this energy back into the zeroth order mode by turning off the gradient, applying the grating again during a time $\tau_s$ but with one arm of the grating shifted by $\pi$. The key point is that, during the refocussing, the first order modes do not cross $k=0$. Turning the gradient on again therefore realizes a similar FIFO read-out as in the $\Lambda$-GEM: the pulses come out without time-reversal but now using only two-level atoms and this purely dispersive diffraction control.

One can even re-arrange the order of pulses. For instance, one can make a sequence of pulses 1-2-3 come out in an order 3-1-2 using a similar combination of diffraction and refocussing techniques. This is demonstrated Fig. 3-b).
First, the third pulse is left to come out after a standard GEM sequence. Then, the FIFO sequence is implemented to reorder the pulses 1 and 2.
The gradient combined with sinusoidal shifts is therefore useful  for engineering quantum memories
and specifically to control pulses of light in atomic ensembles that lack a $\Lambda$ scheme.

Let us now discuss how the above mentioned effects can be demonstrated using current technologies. Cold gases of alkali atoms can be used to store light efficiently using GEM \cite{Hos11} but
due to the short lifetime of their excited states, one shall use the $\Lambda$-GEM \cite{Het08}. The phase grating can then be induced by a pair of circularly polarized lasers tuned to the transition used to Raman couple in order to induce a differential light shift to the ground states. 
With cold atoms as a medium however, the temperature should be low enough so that atomic motion does not wash out the higher order modes of the spin wave. This is especially important when one uses counter-propagating fields where the grating spacing will be on the order of hundred's of nanometers. We indeed find that assuming a grating step of 400 nm and a storage time of 0.1 ms, atomic temperatures of less than 1 $\mu$K are required for the atoms to feel the valleys of the standing wave and thus for the atomic coherences to be diffracted, favouring quasi-copropagating geometries where magneto-optical traps or even warm vapours can be used. Conversely, provided the light grating does not influence atomic motion, this diffraction effect can be used as a very sensitive quantum thermometer for ultracold atoms.
The narrow two-level atomic transitions in rare earth doped crystals \cite{Hed10} are also ideal for the effect to be observed experimentally.
The sinusoidal detuning can be realized simply by using a 
light shift applied directly close to the optical transition and the M\H{o}ssbauer effect prevents the standing wave to induce atomic motion. 
The light fields should only be tuned so that stimulated and spontaneous emission are negligible.
Finally, being able to control the two-level atom coupling to a given mode in rare earth doped materials 
can be useful for turning delay lines into memories without the noise induced by $\pi$-pulses \cite{Bonarota, Afzelius}
thus widening the range of materials available for quantum memory applications.

In conclusion, we present a mechanism akin to diffraction of optical fields or electron motion, where coherent superpositions of spins are diffracted and efficiently read-out.
Applying far-detuned laser pulses along an array of atomic spin coherences is shown to generate an atomic diffraction pattern in momentum space that can be rephased, engineered to control the coupling to light and detected via a gradient echo mapping to optical fields. Besides the proposed coherent pulse sequencer for two-level atom memories, the very high sensitivity of diffraction to atomic temperature has implications for tunable and non-invasive cold atom thermometry.

We would like to acknowledge useful discussions  with B.C. Buchler, P.K. Lam
and financial support from ANR retour post-doctorants SMEQUI (ANR-13-PDOC-0024-01).

\end{document}